\documentclass[11pt]{article}
\usepackage{graphicx}
\usepackage{moriond}

\newcommand{\BABARPubYear}    {01}

\newcommand{\BABARProcNumber} {29}
\newcommand{\SLACPubNumber} {8835}

\input babarsym

\newcommand{\UM}{\mbox{$\,\mathrm{\mu m}$}}
\newcommand{\MM}{\mbox{$\,\mathrm{mm}$}}
\def\BCP {\ensuremath{B_{CP}}\xspace}
\def\BFLAV {\ensuremath{B_{\rm flav}}\xspace}
\def\BCH {\ensuremath{B_{\rm ch}}\xspace}
\def\BTAG {\ensuremath{B_{tag}}\xspace}
\def\deltatreco {\ensuremath{\deltat_{\mathrm rec}}\xspace}

\newcommand{\deltat}{\ensuremath{\Delta t}}

\newcommand{\etaCP}{\ensuremath{\eta_{\CP}}}

\newcommand{\etal}{{\it et al.}}

\long\def\inst#1{\par\nobreak\kern 4pt\nobreak
    {\it #1}\par\vskip 10pt plus 3pt minus 3pt}

\begin{document}
{\pagestyle{empty}

\begin{flushright}
SLAC-PUB-\SLACPubNumber \\
\babar-PROC-\BABARPubYear/\BABARProcNumber \\
May 23, 2001 \\
\end{flushright}

\par\vskip 4cm

\begin{center}
\Large \bf CP VIOLATION, $B$ MIXING AND $B$ LIFETIME RESULTS\\ FROM THE BABAR EXPERIMENT
\end{center}
\bigskip

\begin{center}
\large 
J.\ Beringer\\
Santa Cruz Institute for Particle Physics\\
University of California, Santa Cruz,
1156 High Street,
Santa Cruz, CA 95064\\
beringer@slac.stanford.edu\\[1ex]
(for the \lbabar\ Collaboration)
\end{center}
\bigskip \bigskip

\begin{center}
\large \bf Abstract
\end{center}

The \babar\ detector at the \pep2 asymmetric \BF\ at SLAC collected a
sample of $23\cdot10^6$ \BB pairs in the years 1999 and 2000.  Using
this data sample, we measure the amplitude of the time-dependent
CP-violating asymmetry in neutral $B$ decays to
the CP eigenstates $\jpsi\KS$, $\psitwos\KS$ and $\jpsi\KL$.  We find
a value of $\stwob = 0.34 \pm 0.20 (stat) \pm 0.05 (syst)$.  We also
present preliminary measurements of the $\Bz\Bzb$ oscillation
frequency and of the lifetimes of charged and neutral $B$ mesons.

\vfill
\begin{center}
Invited talk given at the\\
36th Rencontres de Moriond on QCD and Hadronic Interactions \\
17-24 March 2001, Les Arcs, France
\end{center}

\vspace{1.0cm}
\begin{center}
{\em Stanford Linear Accelerator Center, Stanford University, 
Stanford, CA 94309} \\ \vspace{0.1cm}\hrule\vspace{0.1cm}
Work supported in part by Department of Energy contract DE-AC03-76SF00515.
\end{center}


The measurement of CP--violating asymmetries in the time
distribution of decays of neutral $B$ mesons can provide a direct
test of the standard model of electroweak interactions \cite{smtest}
and is the primary goal of the \babar\ experiment at
the \pep2 asymmetric--energy \epem collider at SLAC.
Decays of $\Bz$ and $\Bzb$ mesons into charmonium CP eigenstates
due to $\b \to \ccbar\s$ transitions (e.g.\ \bpsiks)
can be used to measure \stwob (where $\beta$ is one of the interior
angles of the unitarity triangle) with negligible
corrections from strong interactions.

In the years 1999 and 2000, \babar\ \cite{babar} collected a sample
of $23\cdot10^6$ \BB pairs (20.7\invfb on--peak) at \pep2,
where \BB mesons are produced at the $\Upsilon(4S)$ resonance
in collisions of 9.0\gev electrons and 3.1\gev positrons.
The boost $\left< \beta\gamma \right> = 0.56$ along the collision axis ($z$)
resulting from the asymmetric beam energies allows the
determination of the proper decay time difference \deltat\ of
the two $B$ mesons from the measurement of the decay length difference \deltaz,
whose average value is $\left< \beta\gamma \right> c \tau_{B^0} \simeq 260\UM$.

In $\Upsilon(4S)$ decays, \BzBzb pairs are produced in
a $P$--wave state and evolve coherently until one of the
$B$ meson decays. At that time ($\deltat = 0$) the other $B$ meson has
the opposite flavor.
In events where one of the $B$ mesons, \BCP, decays into a charmonium CP eigenstate 
and the other, \BTAG, decays such that its flavor can be
determined, the expected decay--time distribution ${\cal F}_+$ (${\cal F}_-$)
for events where the flavor tag is a \Bz (\Bzb) is given by
\begin{equation}
\label{eq:sin2b}
{\cal F}_\pm(\deltatreco; \Gamma, \deltamd, \mistag, \sin{ 2 \beta } )  
= {\frac{1}{4}}\Gamma {\rm e}^{ - \Gamma \left| \deltat \right| }
\left[ 1 
\mp\etaCP (1-2\mistag)\sin{ 2 \beta } \sin{ \deltamd \deltat } \right]
\otimes  {\cal {R}}(\delta_{\rm t}; \hat {a}) \, .
\end{equation}
The mistag rate \mistag is the probability to wrongly determine
the flavor of \BTAG. The decay--time distribution is
convoluted with a time resolution function
${\cal R}(\delta_{\rm t}=\deltatreco-\deltat; \hat {a})$ 
with parameters $\hat a$ in order to account for the finite resolution of the detector.
$\etaCP$ is the CP eigenvalue
of the final state (-1 for decay modes with \KS, +1 for modes with \KL).

For events where one of the $B$ mesons decays into a fully reconstructed flavor
eigenstate \BFLAV, the decay--time distribution for unmixed (${\cal H}_+$) and
mixed (${\cal H}_-$) signal events is
\begin{equation}
\label{eq:mixing}
{\cal H}_\pm(\deltatreco; \Gamma, \deltamd, \mistag, \hat {a} )  = 
{\frac{1}{4}}\,\Gamma\, {\rm e}^{ - \Gamma \left| \deltat \right| }
\left[  1 \pm (1-2\mistag)\cos{ \deltamd \deltat } \right]
\otimes 
{\cal {R}}( \delta_{\rm t} ; \hat {a} ) \, ,
\end{equation}
where an unmixed event is one where the \BFLAV and \BTAG mesons have opposite flavor.
The mistag rate and resolution function
(the latter is dominated by the reconstruction of the \BTAG vertex)
are the same as for the \BCP sample, and both can be determined from
the \BFLAV sample.

CP candidates are reconstructed in the decay modes
$\jpsi\KS$, $\psitwos\KS$, and $\jpsi\KL$
and are required to have a difference $\Delta E$ between the energy
of the \BCP candidate and the beam energy in the center--of--mass frame
of less than 3 standard deviations from zero. In addition, modes with
a \KS must have a beam--energy substituted mass
$\mes=\sqrt{{(E^{\rm cm}_{\rm beam})^2}-(p_B^{\rm cm})^2} > 5.2\gevcc$
($\mes > 5.27\gevcc$ for candidates counted as signal).
The distributions of \mes and $\Delta E$ for the \BCP candidates are shown
in figure \ref{fig:bcp}.
A sample of $B$ decays, \BFLAV, reconstructed in the flavor eigenstate modes \footnote{
Throughout this paper, flavor--eigenstate decay modes imply also their charge conjugate.}
$D^{(*)-}h^+(h^+=\pi^+,\rho^+,a_1^+$) and  $\jpsi K^{*0}\ (K^{*0}\to K^+\pi^-)$
is used to measure the \Bz lifetime and \deltamd.
A sample of charged $B$ decays, \BCH, in the final states $\jpsi K^{(*)+}$, $\psitwos K^+$  
and $\Dbar^{(*)0}\pi^+$ is used to measure the \Bu lifetime and serves
as a control sample. Yields and purities for events with a flavor tag are summarized in table~1.

\begin{figure}[t]
\begin{center}
\begin{minipage}{6.2cm}
\includegraphics[width=5.8cm]{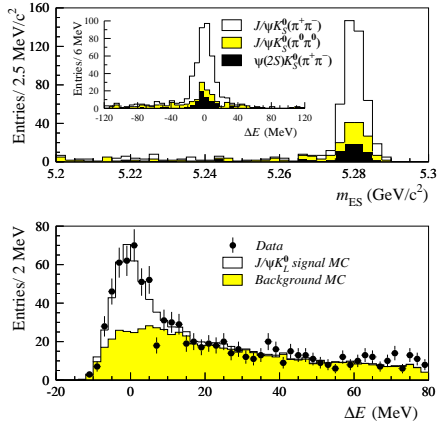}
\vspace{-0.3cm}
\caption{\label{fig:bcp}\mes\ and $\Delta E$ distribution for \BCP\ candidates with \KS (top) and \KL (bottom).}
\end{minipage}
\hfill
\begin{minipage}{8.8cm}
\begin{center}
\small
\begin{tabular}{|lccr|}
\hline
Sample                                   & $N_{\rm tag}$& Purity (\%)    &  \stwob\hspace{2ex}  \\ \hline \hline
Full \CP\ sample                        & $529$        & $69\pm2$       &  0.34 $\pm$ 0.20  \\ \hline
$\jpsi \KL$                              & $256$        & $39\pm6$       &  0.87 $\pm$ 0.51  \\ \hline
$\jpsi\KS$, $\psitwos\KS$                & $273$        & $96\pm1$       &  0.25 $\pm$ 0.22  \\ 
- {\tt Lepton} tags  & $34$         &  $99\pm2$      &  0.07 $\pm$ 0.43  \\ 
- {\tt Kaon} tags    & $156$        &  $96\pm2$      &  0.40 $\pm$ 0.29  \\ 
- {\tt NT1} tags     & $28$         &  $97\pm3$      &  -0.03 $\pm$ 0.67 \\ 
- {\tt NT2} tags     & $55$         &  $96\pm3$      &  0.09 $\pm$ 0.76  \\ 
- \Bz\ tags                           & $141$        &  $96\pm2$      &  0.24 $\pm$ 0.31  \\ 
- \Bzb\ tags                          & $132$        &  $97\pm2$      &  0.25 $\pm$ 0.30  \\ 
\hline
$B_{\rm flav}$ sample                    & $4637$       & $86\pm1$       &  0.03 $\pm$ 0.05  \\
\hline 
Charged $B$ sample                       & $5165$       & $90\pm1$       &  0.02 $\pm$ 0.05  \\ \hline
\end{tabular}
\begin{minipage}{8.8cm}
\vspace*{0.3cm}
\footnotesize
Table 1: Number of tagged events, signal purity, and result of fitting for CP
asymmetries in the full \BCP sample, in various subsamples, and in the
\BFLAV and charged $B$ samples.
\end{minipage}
\end{center}
\end{minipage}
\end{center}
\end{figure}

The vertex of the other $B$ in the event (\BTAG) is determined by fitting the tracks
not belonging to the reconstructed \BCP, \BFLAV or \BCH to a common vertex. Tracks from $\gamma$ conversion
are removed and reconstructed \KS and $\Lambda$ candidates are used as input to the
fit in place of their daughters. Tracks with a large ($>6$) $\chi^2$ contribution
are removed to reduce the bias from charm decays. We require
$\sigma(\deltaz) < 400 \UM$ and $\left| \deltaz \right| < 3\MM$.
The average resolution of $\deltaz$ is 190\UM.

To a very good approximation $\deltat \approx \deltaz / {\rm c} \left< \beta \gamma \right>$, but
event--by--event corrections are made for the direction of the $B$ with respect
to the $z$ direction in the $\Upsilon(4S)$ frame.
The resolution function ${\cal R}(\delta_{\rm t}; \hat {a})$ is parameterized either as the sum
of three Gaussian distributions (\stwob and \deltamd measurements) or as the sum of a
zero--mean Gaussian distribution and its convolution with a decaying exponential
(lifetime measurements). In both cases, the event--by--event errors calculated by the
vertex fits are used to scale some of the contributions to
${\cal R}(\delta_{\rm t}; \hat {a})$.
From an unbinned maximum likelihood fit to the \deltatreco\ distribution in the \BFLAV and \BCH samples,
including also events without a flavor tag, we obtain the preliminary results
\begin{eqnarray}
\tau_{\Bz} & = & 1.546 \pm 0.032\,{\rm (stat)} \pm 0.022\,{\rm (syst)}\,{\rm ps}\\
\tau_{\Bu} & = & 1.673 \pm 0.032\,{\rm (stat)} \pm 0.022\,{\rm (syst)}\,{\rm ps}\\
\tau_{\Bu}/\tau_{\Bz} & = & 1.082 \pm 0.026\,{\rm (stat)} \pm 0.011\,{\rm (syst)} \, .
\end{eqnarray}
The fit takes into account contributions from signal, background and outliers. The
probability of each candidate to be signal   is determined from the \mes distribution.
An empirical description based on \mes sidebands and including both prompt
and lifetime components is used to describe the \deltat\ shape of the combinatoric background
from other B decays and from continuum events.

In order to determine the \BTAG's flavor tag
for the \stwob and \deltamd measurements,
we use the flavor information carried by the charge of high momentum
leptons ($e$, $\mu$) from semileptonic $B$ decays,
of kaons from $b \to c \to s$ transitions, of soft
pions from $D^*$ decays and of high momentum charged particles
not coming from the reconstructed \BCP or \BFLAV candidate.
Each event is assigned to one of four hierarchical mutually exclusive
tagging categories (or else not assigned a flavor tag). The {\tt Lepton}
and {\tt Kaon} categories are characterized by the presence of
an electron or muon with a center--of--mass momentum $p_e^* > 1.0\gevc$
or $p_\mu^* > 1.1\gevc$,
and of one or more kaons with a non--zero charge sum, respectively.
The remaining two categories, {\tt NT1} and {\tt NT2}, are based on the
output of a neural network algorithm whose performance relies
primarily on soft pions and on recovering isolated electrons and
muons from semileptonic $B$ decays.
The tagging performance measured on data is shown in table~2.

\begin{figure}[t]
\begin{center}
\begin{minipage}{9.6cm}
\hspace{0.3cm}
\label{tab:tagging}
\begin{center}
\small
\begin{tabular}{|lrrrr|}
\hline
Category     & $\varepsilon$ (\%)\hspace{2ex} & $\mistag$ (\%) \hspace{1ex} & $\Delta\mistag$ (\%)\ & $Q$ (\%) \hspace{0.5ex} \\ \hline
{\tt Lepton} & $10.9\pm0.4$ & $11.6\pm2.0$ & $3.1\pm3.1$ &   $6.4\pm0.7$  \\ 
{\tt Kaon}   & $36.5\pm0.7$ & $17.1\pm1.3$ & $-1.9\pm1.9$ &  $15.8\pm1.3$  \\ 
{\tt NT1}    & $ 7.7\pm0.4$ & $21.2\pm2.9$ & $7.8\pm4.2$  &   $2.6\pm0.5$  \\ 
{\tt NT2}    & $13.7\pm0.5$ & $31.7\pm2.6$ & $-4.7\pm3.5$ &   $1.8\pm0.5$  \\ \hline
All          & $68.9\pm1.0$ &              &          &  $26.7\pm1.6$  \\ \hline
\end{tabular}
\end{center}
\begin{minipage}{9.6cm}
\vspace*{0.3cm}
\footnotesize
Table 2: Average mistag fractions $\mistag_i$ and mistag differences $\Delta\mistag_i=\mistag_i(\Bz)-\mistag_i(\Bzb)$
extracted for each tagging category $i$ from the maximum-likelihood fit to the time distribution in the
\BFLAV+\BCP sample. The figure of merit for tagging is 
$Q_i = \eps_i (1-2\mistag_i)^2$, where $\eps_i$ 
is the fraction of events
in the $i^{th}$ category. 
The  statistical error on \stwob is proportional to $1/\sqrt{Q}$, where $Q=\sum Q_i$.
\end{minipage}
\end{minipage}
\hspace{0.1cm}
\begin{minipage}{6cm}
\begin{center}
\includegraphics[width=5.5cm,height=4cm]{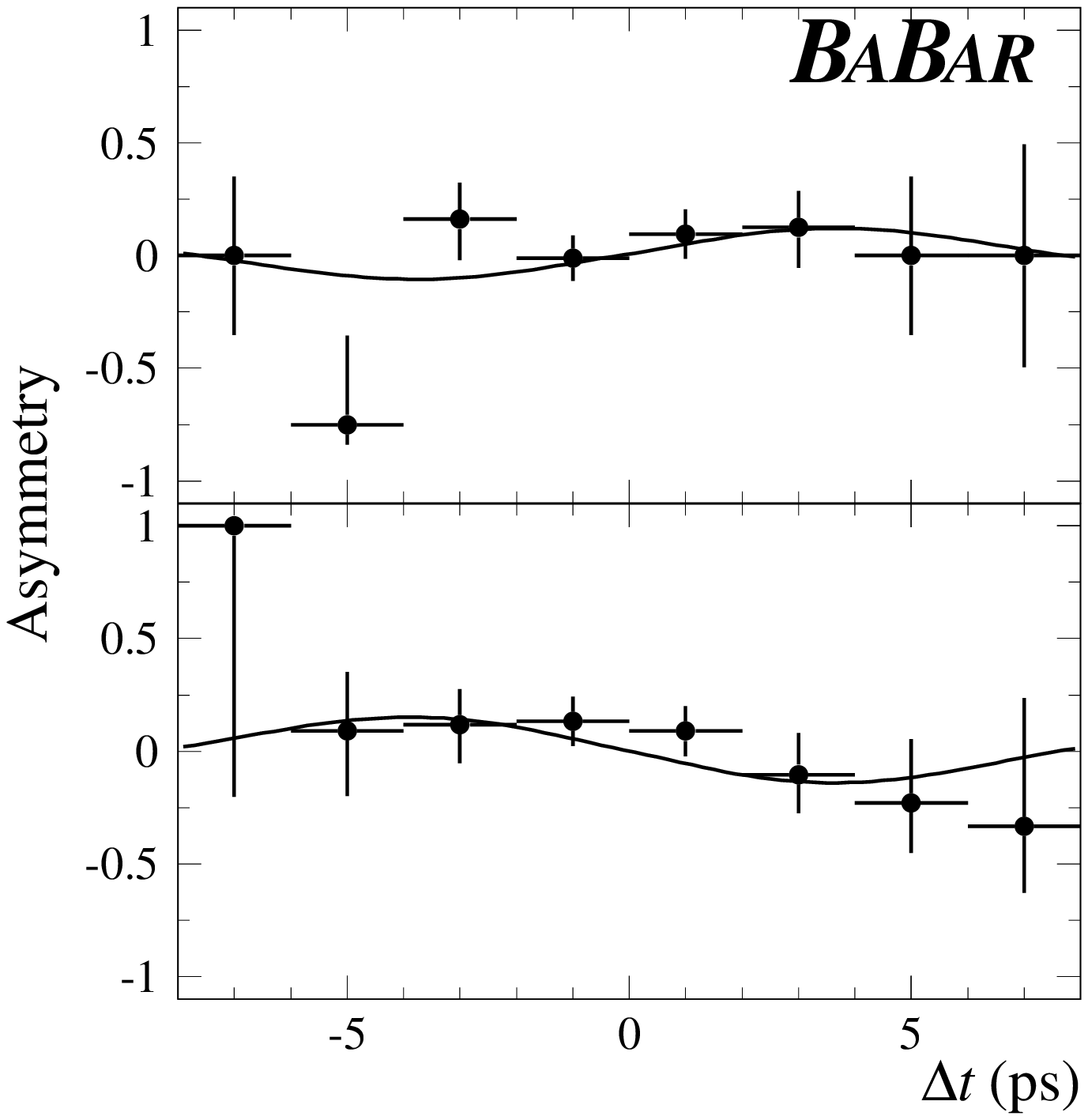}
\vspace{-2ex}
\caption{\label{sin2b-asym} Raw asymmetry in the number of \Bz and \Bzb tags for
\jpsi\KS, \psitwos\KS (top)
and \jpsi\KL modes (bottom). The solid curve
is the \stwob fit to these samples.}
\end{center}
\end{minipage}
\end{center}
\end{figure}

Based on the flavor of the reconstructed \BFLAV candidate and the flavor tag,
events in the \BFLAV\ sample are classified as mixed or unmixed.
The time--dependent mixing asymmetry
$A(\deltatreco) = (N_{unmixed}-N_{mixed})/(N_{unmixed}+N_{mixed})$ is shown in figure~\ref{mixing-asym}.
\deltamd is determined from an
unbinned maximum likelihood fit in which the mistag fractions $\mistag_i$ and
$\Delta\mistag_i$ (8 parameters), signal resolution function parameters (9 parameters)
and the fractions, lifetimes, dilutions and resolution function parameters of different background components
(16 parameters) are determined simultaneously with \deltamd. The correlation
between \deltamd and $\mistag_i$ is small because
the latter are determined by events at low values of \deltat\ where
the mixing probability is small.

An alternative method for measuring \deltamd is to use inclusively reconstructed dilepton events,
i.e.\ events where both $B$ mesons decay semileptonically and the flavor
of each $B$ is given by the charge of the high momentum electron or muon
produced in its decay. Because of the relatively large semileptonic branching
ratio and the high lepton identification efficiency, this method is statistically
more powerful. The non--negligible backgrounds
due to leptons from charm decays are minimized with a neural network
technique which uses the lepton momenta and opening angle, and the total and
missing energy as input.
The resulting dilepton sample has about equal contributions from neutral and charged $B$ mesons,
but the former can be enhanced by an
inclusive reconstruction of $\Bzb \rightarrow D^{*+} \ell^- \nu$ decays.
The mixing asymmetry obtained with this technique is shown in
figure~\ref{mixing-asym}.

The preliminary results obtained for \deltamd\ with the two methods are:
\begin{eqnarray}
\deltamd   & = 0.519 \pm 0.020\,{\rm (stat)} \pm 0.016\,{\rm (syst)}\, \hbar \, {\rm ps}^{-1} \hspace{1cm} & \mathrm{(\BFLAV\ sample)} \\
\deltamd   & = 0.499 \pm 0.010\,{\rm (stat)} \pm 0.012\,{\rm (syst)}\, \hbar \, {\rm ps}^{-1} \hspace{1cm} & \mathrm{(dilepton\ sample)}
\end{eqnarray}

The \stwob measurement is made with an unbinned maximum likelihood fit
to the \deltatreco\ distribution of the tagged candidates from
the combined \BCP and \BFLAV samples with parameters similar to
the ones used in the fit for \deltamd. The values of the \Bz lifetime and
\deltamd are fixed to their world average \cite{pdg}.
We find a value of\, \cite{sin2bprl}
\begin{equation}
\stwob = 0.34 \pm 0.20\,{\rm (stat)} \pm 0.05\,{\rm (syst)} \, .
\end{equation}
The raw asymmetry in the number of \Bz and \Bzb tags as a function of \deltatreco\
is shown in figure~\ref{sin2b-asym} and has, as expected from eq.~\ref{eq:sin2b},
the opposite sign for CP even and CP odd modes.

The determination of the mistag rates and \deltat\
resolution function function is dominated by the
high statistics \BFLAV sample.
The largest correlation between \stwob and any linear
combination of the other 34 free parameters is 0.076.
The dominant sources of systematic error are the
assumed parameterization of the \deltat\ resolution
function, due in part to residual uncertainties in
the alignment of the silicon vertex tracker, and
uncertainties in the level, composition and CP asymmetry
of the background in the selected CP events.
The large \BCP sample allows a number of consistency checks, including
fits to subsamples of the data by decay mode, tagging
category and flavor of the \BTAG as shown in table~1.
No statistically significant asymmetry is found in
fits to control samples where no asymmetry is expected.

The measured value of \stwob is consistent with the range
implied by measurements and theoretical estimates of the
magnitudes of CKM matrix elements \cite{gilman}. It
is also consistent with no CP asymmetry at the 1.7$\sigma$
level.

\begin{figure}[t]
\begin{center}
\includegraphics[height=5.2cm]{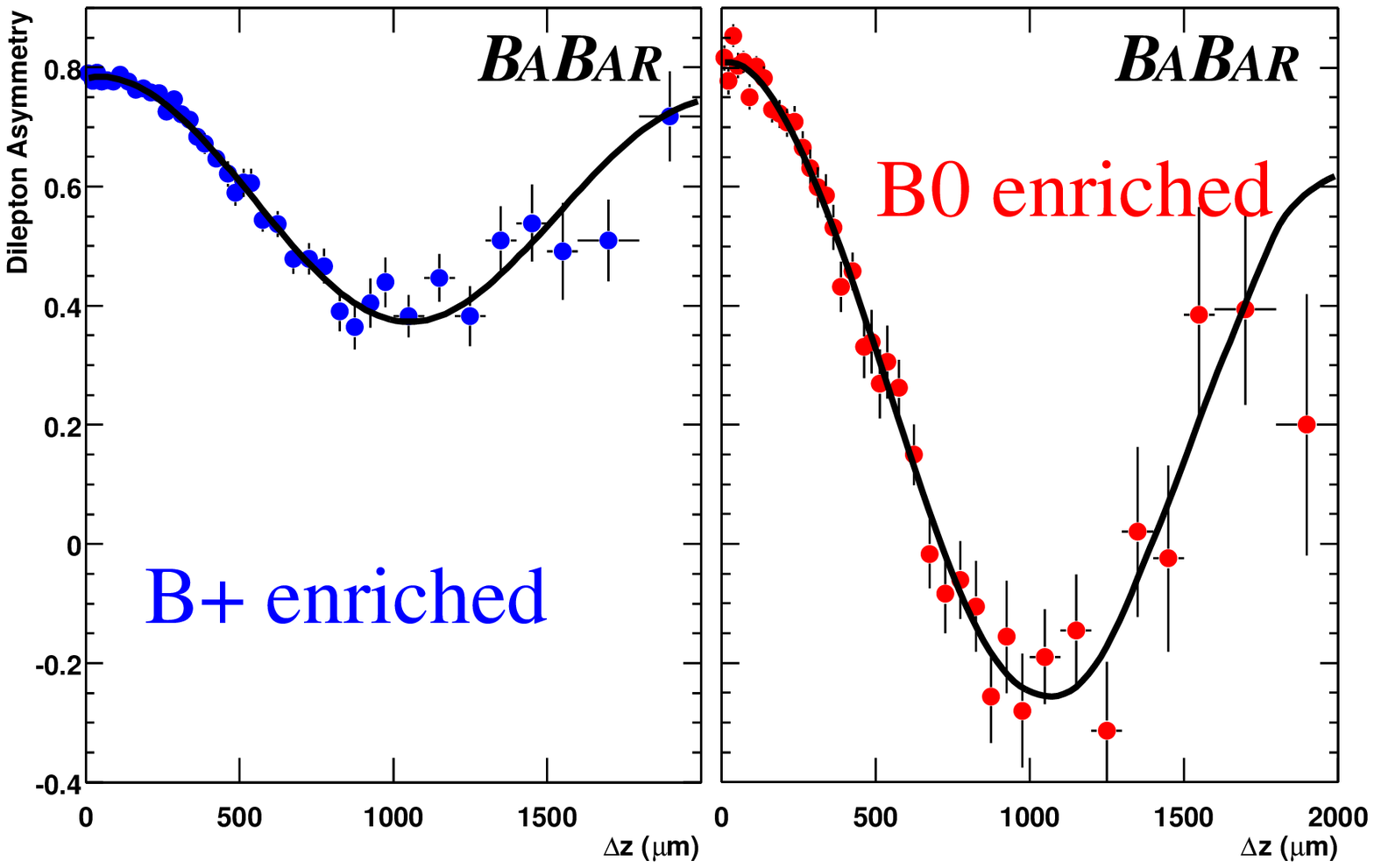}
\hspace{0.5cm}
\includegraphics[height=5.3cm]{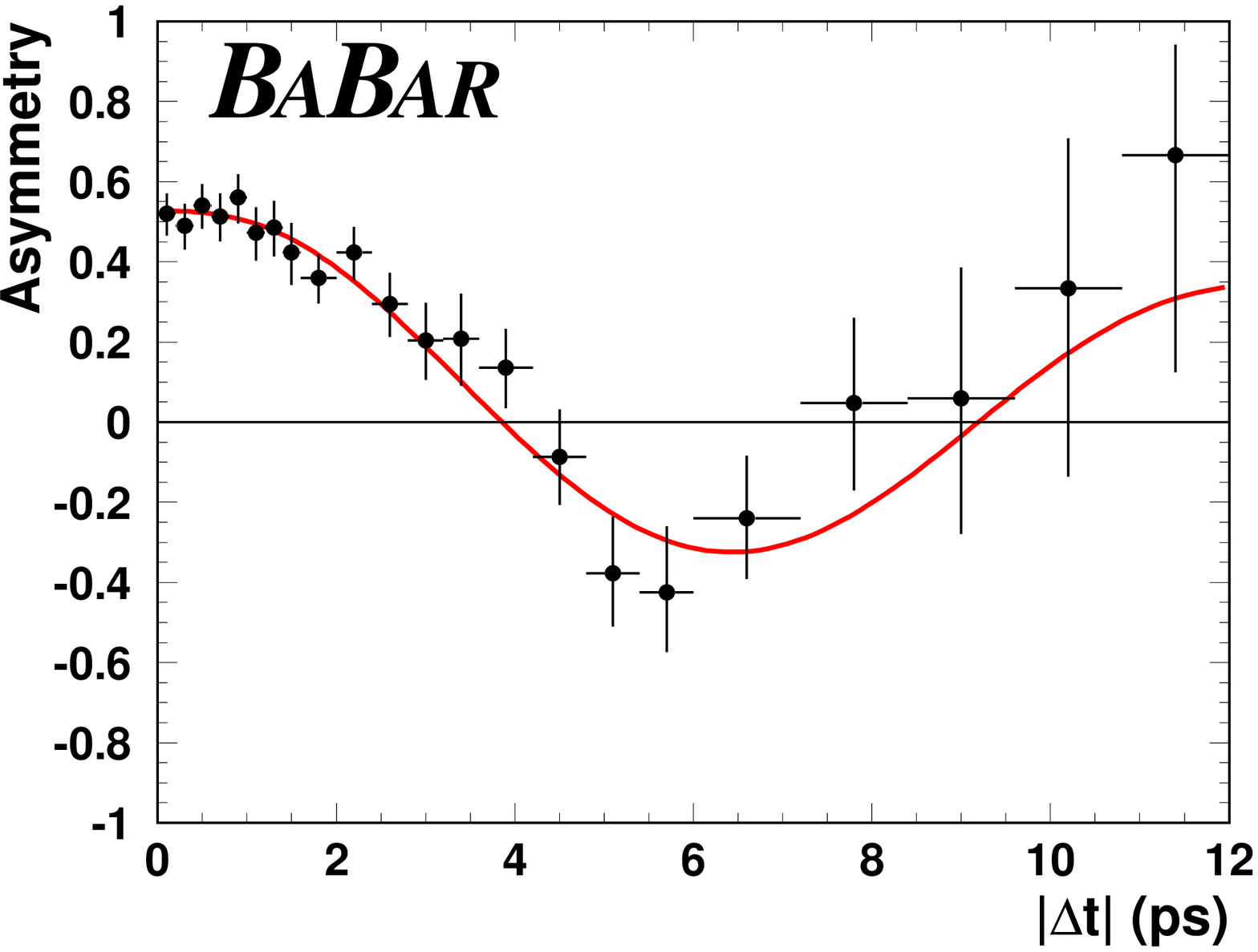}
\vspace{-0.5cm}
\caption[]{\label{mixing-asym} Mixing asymmetry $A(\deltatreco)$ in the dilepton (left) and the \BFLAV (right) sample (preliminary).}
\end{center}
\end{figure}

\end{document}